\documentclass[aip,apl,reprint,letterpaper,floatfix,nobalancelastpage]{revtex4-1}
\usepackage{graphicx}
\usepackage{amsmath} 
\usepackage{amssymb}

\begin{document}

\title{Improved superconducting qubit coherence using titanium nitride}

\author{J.~Chang}
\altaffiliation{Contributed equally to this work}
\affiliation{IBM T.J. Watson Research Center, Yorktown Heights, NY 10598, USA}
\author{M.~R.~Vissers}
\altaffiliation{Contributed equally to this work}
\affiliation{National Institute of Standards and Technology, Boulder, Colorado 80305, USA}
\author{A.~D.~C\'orcoles}
\affiliation{IBM T.J. Watson Research Center, Yorktown Heights, NY 10598, USA}
\author{M.~Sandberg}
\affiliation{National Institute of Standards and Technology, Boulder, Colorado 80305, USA}
\author{J.~Gao}
\affiliation{National Institute of Standards and Technology, Boulder, Colorado 80305, USA}
\author{David W.~Abraham}
\affiliation{IBM T.J. Watson Research Center, Yorktown Heights, NY 10598, USA}
\author{Jerry M.~Chow}
\affiliation{IBM T.J. Watson Research Center, Yorktown Heights, NY 10598, USA}
\author{Jay M.~Gambetta}
\affiliation{IBM T.J. Watson Research Center, Yorktown Heights, NY 10598, USA}
\author{M.~B.~Rothwell}
\affiliation{IBM T.J. Watson Research Center, Yorktown Heights, NY 10598, USA}
\author{G.~A.~Keefe}
\affiliation{IBM T.J. Watson Research Center, Yorktown Heights, NY 10598, USA}
\author{Matthias Steffen}
\affiliation{IBM T.J. Watson Research Center, Yorktown Heights, NY 10598, USA}
\author{D.~P.~Pappas}
\affiliation{National Institute of Standards and Technology, Boulder, Colorado 80305, USA}

\begin{abstract}
We demonstrate enhanced relaxation and dephasing times of transmon qubits, up  to $\sim 60$ $\mu$s by fabricating the interdigitated shunting capacitors using titanium nitride (TiN). Compared to lift-off aluminum deposited simultaneously with the Josephson junction, this represents as much as a six-fold improvement and provides evidence that previous planar transmon coherence times are limited by surface losses from two-level system (TLS) defects residing at or near interfaces. Concurrently, we observe an anomalous temperature dependent frequency shift of TiN resonators which is inconsistent with the predicted TLS model.
\end{abstract}

\maketitle

Long coherence times compared to logic gate times are necessary for building a fault tolerant quantum computer. In the case of superconducting qubits, coherence times have dramatically improved since the first demonstration \cite{Nakamura99}, by a factor of $10^{3}$ planar circuits \cite{Chow12} and $10^{4}$ using the 3D architecture \cite{Paik11,Rigetti12}. Most of the improvements are attributable to a number of design changes including removing dissipation from the chip by coupling qubits to resonators (cQED architecture) \cite{Wallraff04}, reducing the impact of charge noise \cite{Koch07}, and varying the geometry of the qubit shunting capacitor \cite{Chow12}. The natural question that arises is how further improvements will be possible, especially for the transmon qubit \cite{Koch07}, which has become a popular superconducting qubit design choice in the community.

To answer this question it is instructive to formulate a hypothesis of what limits currently observed coherence times of transmons the majority of which are fabricated using lift-off aluminum deposited simultaneously with the Josephson junction. We believe the physical origin of dissipation is likely related to dielectric loss of amorphous materials \cite{Martinis05}. Clever design improvements have reduced the impact of this loss mechanism but have not completely solved it. Specifically, two-level system (TLS) dielectric loss near interfaces appears to play a crucial role \cite{Wenner11}. Experimental evidence to support this claim comes from (1) various design improvements and (2) materials improvements. 

First, quality factors are enhanced by geometric variations of CPW resonators~\cite{Gao08,Wang09a,Geerlings12}, qubit coherence times improve by increasing the size of the shunting capacitors \cite{Chow12}, as well as by employing the 3D architecture \cite{Paik11,Rigetti12}. In fact, the recent advent of the 3D architecture highlights that ultra-small ($\sim 100$~nm) Josephson junctions likely do not play a key role in dissipation at the moment, and neither does bulk substrate loss (in the case of sapphire), leaving behind dielectric loss at interfaces ("surface losses") as a potential leading contributor to decoherence. 

Second, it was shown that quality factors of resonators using aluminum on sapphire can be improved by careful surface treatment prior to the deposition of the aluminum~\cite{Megrant12}. Similarly, titanium nitride (TiN) was also employed to improve the quality factors of resonators~\cite{Vissers10}, as well as for cQED devices demonstrating $T_1 \sim T_2 \sim 11$~$\mu$s \cite{Sandberg13}. However, these circuits used a large capacitor pad geometry for the transmons and were found to be predominantly limited by the Purcell effect and radiation loss. 

Here we use  TiN to fabricate more traditional interdigitated shunting capacitor (IDC) based transmon qubits and show substantial improvements in qubit coherence times are possible, while leaving the qubit design nominally unaltered. We measure coherence times as long as $T_1 \sim T_2 \sim 55-60$~$\mu$s compared to only $T_1 \sim T_2 \sim 18$~$\mu$s using lift-off aluminum deposited at the same time as the Josephson junction. Rather surprisingly, we also observe that the predicted frequency shift of the readout resonator versus temperature does not follow the predicted model from TLS, indicating that observed frequency shifts may not be an accurate indicator of TLS dissipation.

\begin{figure}[htbp!]
		 \centering
	\includegraphics[width=0.47\textwidth]{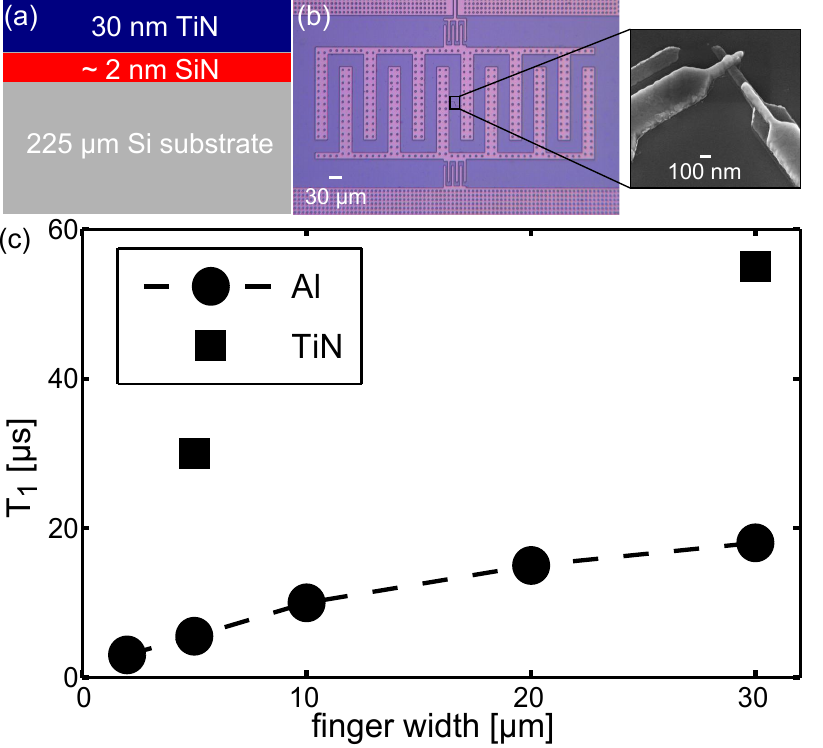}
		 \caption{Description of the qubit and summary of coherence times. (a) The fabrication process involves a thin seed layer of SiN on top of Silicon. (b) Micrograph of the transmon, and (c) zoom in of the Josephson junction. (d) $T_1$ versus finger width dimension, keeping the finger width and separation constant for an interdigitated capacitor of total capacitance $C \sim 100-150$~fF. Coherence times of qubits using aluminum (circles) for the capacitor structure improves with increasing finger width. The dashed line is a guide to the eye. A dramatic improvement is observed using TiN (squares). All qubit frequencies were in the range $3-5$~GHz and the multi-mode Purcell limit is predicted to be far away from observed coherence times ($>4$~ms).}
		 \label{fig:fig1}
\end{figure}

The IDCs were fabricated from TiN grown on silicon substrates. The substrates were high resistivity ($\rho$ $>$ 20~k$\Omega$-cm) intrinsic Si wafers that are prepared using a HF etch. This acts to remove the native oxide from the surface, leaving a hydrogen terminated surface which remains stable long enough to be transferred to the deposition system and pumped down. Next, the surface is heated to 500$^\circ$C.  Then a Ti target is reactively DC sputtered using a mixture of Ar and N$_2$ gases. While the sputter source is stabilizing with the shutter still closed, the resulting Ar:N$_2$ plasma forms a thin, (d$\approx$2~nm), amorphous SiN layer on the substrate as is shown in Figure 1a.  This buffer layer allows both a low-loss interface as well as the growth of a (200)-textured TiN film~\cite{Vissers10}. A 30 nm thick film of the TiN was then deposited using reactive sputtering from a pure Ti target in the Ar:N$_2$ at the same temperature. A 4W RF 100V DC self-bias was also added to the substrate during growth.

Afterwards, the qubit is fabricated as follows. Using a shadow mask, niobium regions are deposited in non-device areas for alignment purposes. The wafer is then patterned with contact lithography in i-line resist and a timed Cl$_2$/BCl$_3$ reactive ion-etch is conducted at low pressure and high bias in an inductively coupled plasma tool. The etch leads to a recess in the silicon substrate of 125 nm. Immediately after etching, the remaining resist is stripped through manual agitation in an acetone bath, and chlorine (Cl) residue removed with a room temperature water bath.  Previous work \cite{Sandberg12} has shown a degradation of quality factor in samples processed with a Cl-based etch. Although our process uses Cl we believe the dechlorination step helps prevent the previously observed quality factor degradation. We have not yet explored F-based etches. The only component not made out of TiN is the Josephson junction which is fabricated using shadow evaporation techniques \cite{Steffen10}. Contact is made to the TiN layers by introducing an ion mill clean step to remove surface oxide.

The qubit structure for a 30 $\mu$m shunt capacitor widths and gaps is shown in Fig.~\ref{fig:fig1}(b) along with a zoom-in of the Josephson junction in Fig.~\ref{fig:fig1}(c). All qubits are weakly coupled to a CPW resonator (coupling strength $g \sim~$40-50~MHz and external quality factor $Q \sim 25,000 - 30,000$), which is $6.14$~mm long and supports a $\lambda/2$ resonance at approximately $6.5$~GHz, consistent a kinetic inductance fraction of approximately $\alpha$ $\sim$ $0.54$. Using appropriate microwave shielding methods \cite{Corcoles11}, qubits are measured employing standard cQED measurements techniques \cite{Wallraff04} at either low or high microwave power \cite{Reed10}.

\begin{table}
\begin{ruledtabular}
\begin{tabular}{|c|c|c|c|c|c|c|}
Name & Pitch [$\mu$m] & $f_{\mathrm{qubit}}$ [GHz] & C [fF] & $I_0$ [nA] & $T_1$ [$\mu s$] & $T_2^*$ [$\mu s$] \\ 
\hline
A & 30 & 2.989 	& 161 	    & 19.5	& 53 & 58 \\
B & 30 & 2.954  & 163       & 19.2  & 55 & 56 \\ 
C & 5  & 3.235  & 120       & 17    & 26 & 21 \\
D & 5  & 3.310  & 127       & 19    & 26 & 2$^*$
\end{tabular}
\end{ruledtabular}
\caption{\label{table:1}Measured and extracted qubit parameters. Sample (D) was connected using poorly thermalized feed lines, likely causing reduced $T_2$ times.}
\label{table:1}
\end{table}

Qubit energy relaxation times ($T_1$) for four measured qubits are substantially improved for TiN compared with lift-off aluminum as shown in Fig.~\ref{fig:fig1}(d) and Table \ref{table:1}. For a 5~(30)~$\mu$m finger width we observe a $6\times$ ($3\times$) improvement with a maximum observed coherence time of almost $60$~$\mu$s, corresponding to a quality factor $Q \sim 10^6$, consistent with that observed for resonators \cite{Vissers10}. The variation of $T_1$ dependent on finger width provides evidence that surface losses are the most dominant decoherence contributors. Dephasing times are also greatly improved, although we observe occasional frequency jumps, a phenomenon that has been reported by other groups \cite{Paik11} as well. Average energy relaxation decay and Ramsey fringes are shown in Fig.~2 along with their distribution over an 18 hour time span.

The observation of improved coherence times is consistent with a reduced TLS surface density \cite{Gao08,Wenner11}, and may be attributable to the silicon nitride interface formed during the Ar:N$_2$ plasma surface treatment prior to the deposition of the TiN \cite{Vissers10}. What sets the limits on the observed coherence times as well as the exact functional dependence on finger width is not clear. One possibility is that TLS surface loss is still dominant, and our experiment enables one in-situ test that may provide some insight: Measuring the temperature dependent frequency shift of a TiN resonator.

\begin{figure}[t!]
		 \centering
	\includegraphics[width=0.5\textwidth]{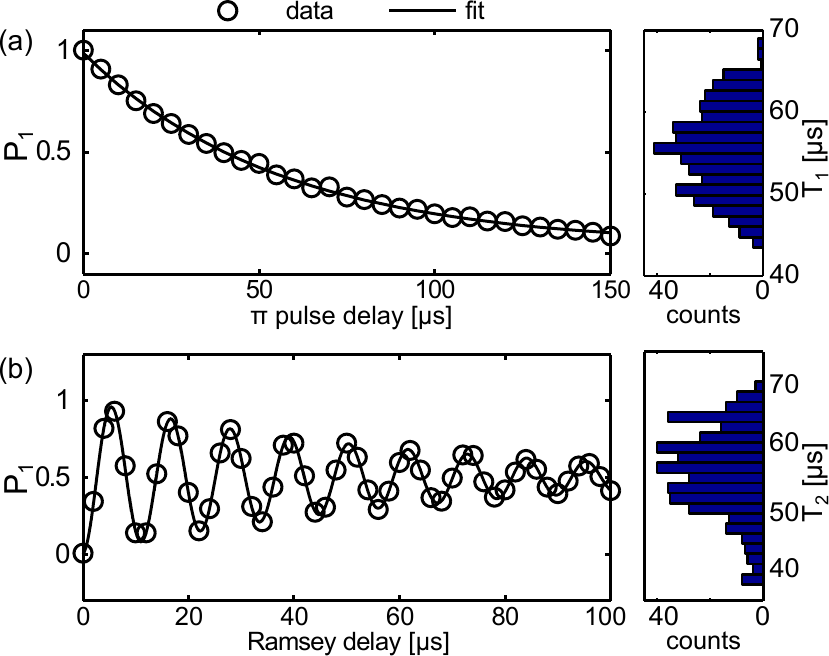}
		 \caption{Representative plots for (a) energy relaxation and (b) Ramsey fringes which give $T_1$ and $T_2$ respectively (sample B) together with a histogram for individually fitted traces run over the course of 18 hours.}
		 \label{fig:fig2}
\end{figure}

From the measured qubit quality factor $Q \sim 0.5$~x~$10^6$ (samples C and D), and assuming (1) the same loss is dominant for a TiN resonator and (2) loss is dominated by TLS dissipation, we can predict a temperature dependent frequency shift for the resonator \cite{Gao08} and compare it with data. Implementing this experiment is possible by using the readout resonator ($f = 6.57$~GHz, $Q \sim 30,000$) which is also made out of TiN, and has similar dimensions (6~$\mu$m center line, 3~$\mu$m gap) as the qubit samples C and D, and should therefore have a similar intrinsic quality factor based on the assumptions above. In order to fully decouple the resonator from the qubit the bare cavity peak is measured~\cite{Reed10}.

\begin{figure}[t]
		 \centering		 		
		 \includegraphics[width=0.47\textwidth]{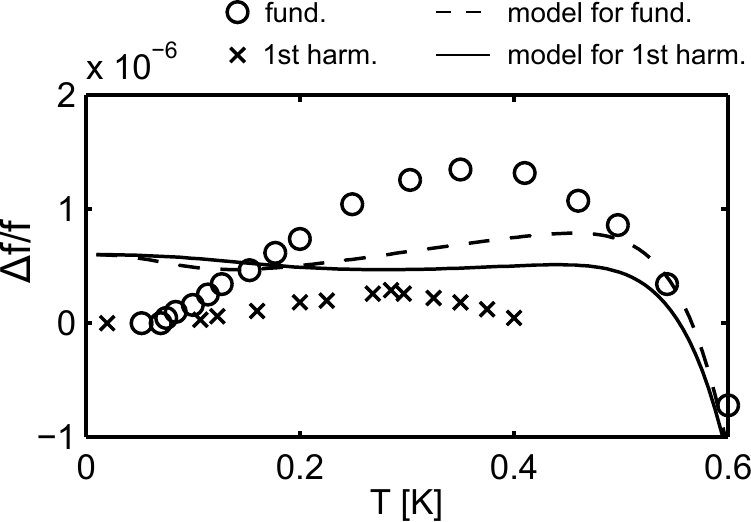}
		 \caption{Frequency shift versus temperature. For the fundamental resonance at $6.57$~GHz the data (open circles) shows a much more pronounced positive shift than predicted (dashed line) assuming $Q_{TLS} \sim 0.5$~x~$10^6$. A similar behavior is found for the first harmonic at $13.14$~GHz (crosses) but the discrepancy compared the prediction (solid line) is not as large. The overall discrepancy is reduced by introducing lower $Q_{TLS}$ in the model which would not be consistent with measured qubit coherence times and assuming the resonator is limited by the same TLS loss. Furthermore, for both the fundamental resonance and first harmonic the predicted local minimum in frequency shift is not observed. Simulations assume $T_c=4.5~K$ and a kinetic inductance fraction of 0.54 which effect the negative frequency trend at elevated temperature. The simulation traces are offset by $\sim$6~x~$10^{-7}$for clarity.}
		 \label{fig:fig3}
\end{figure}

The resulting temperature ($T$) dependent frequency shift ($\Delta f = f(T)-f(T_0)$, where $T_0\approx 15$~mK is the base temperature of the dilution refrigerator) is plotted in Fig.~\ref{fig:fig3} for the fundamental resonance and first harmonic. Although a clear positive shift is present, it does not correlate with the expected shift based on a TLS-limited internal quality factor of $Q \sim 0.5$~x~$10^6$, consistent with previous observations \cite{Gao13}. The measured frequency shift magnitude is much larger than expected, and the predicted local minimum is unobservable. 

While a standard TLS fit to the positive slope in $\Delta f/f$ data would normally indicate a large loss, $\sim 10^{-5}$ or more, the fact that we observe very long $T_1$ in qubits using TiN shows that the total loss must be $10^{-6}$ or less. Therefore, the measured $\Delta f/f$ is in excess of the TLS effect and we suspect this excess positive frequency shift may be related to the anomalous electrodynamical response of TiN observed earlier in kinetic inductance photon detectors~\cite{Gao12,Driessen12}. Other explanations include effects from the nonlinear kinetic inductance contribution at elevated microwave powers. Independent of the root cause, our results indicate that relying on a temperature dependent frequency shift of resonators is not necessarily an accurate predictor of TLS contribution. 

Clearly, resonator experiments using TiN exhibit novel features that need further investigation. Nonetheless, the material in conjunction with appropriate growth conditions can be used to fabricate superconducting transmon qubits with coherence times up to $T_1 \sim T_2 \sim 60~\mu s$, much improved over those obtained with lift-off aluminum \cite{commenth}. What sets the limits to the observed coherence times using TiN is not well understood but potential sources of decoherence include radiation or bulk dielectric loss. The results also suggest that conventional transmons with lift-off Aluminum are limited by surface loss. We believe many other materials may be possible including NbN, NbTiN, and Re to help push coherence times even further by reducing surface loss, because the Josephson junction is independent from the metal deposition of the shunting capacitor and its associated annealing procedures.

The authors would like to thank E. Duch, R. Martin, and E. O'Sullivan for their contributions to this work, and Chris Lirakis, Jim Rozen and Jack Rohrs for their support. The authors acknowledge useful comments on the manuscript from S. Poletto, E. Lucero, N. Masluk, and D. McClure. Portions of this work were completed in the IBM T.J. Watson Microelectronics Research Laboratory. This work was supported by DARPA and the NIST Quantum Information Program. The views and conclusions contained in this document are those of the authors and should not be interpreted as representing the official policies, either expressly or implied, of the Defense Advanced Research Projects Agency.


\begin{thebibliography}{22}
\bibitem [{com()}]{commenth}%
  \BibitemOpen
  \href@noop {} {}\bibinfo {note} {The substrates used for the aluminum and TiN
  based qubits are not identical, with the latter ones fabricted on higher
  resistivity Silicon substrates (1 k$\Omega$-cm versus $>$20 k$\Omega$-cm).
  The coupling capacitors to the measurement resonators also had a different
  pitch (2 $\mu$m versus 5 $\mu$m). However, we believe none of these
  differences contribute strongly to our results or conclusions.}\BibitemShut
  {Stop}%
\end{thebibliography}
\end{document}